\documentclass[ALICE,manyauthors]{cernphprep}

\usepackage{hyperref}
\usepackage{graphicx}  % needed for figures
\usepackage{dcolumn}   % needed for some tables
\usepackage{bm}        % for math
\usepackage{amssymb}   % for math
\usepackage{amsfonts}
\usepackage{graphics}
\usepackage{grffile}   % to handle dots in graphics file names
\usepackage{epsfig}
\usepackage{units}
\usepackage[usenames]{color}
\usepackage[normalem]{ulem} % for strikethroughs (\sout{})
\usepackage{color}
\usepackage[utf8]{inputenc}
\usepackage[T1]{fontenc}
\usepackage{subfigure}
\usepackage{lineno}
\usepackage{cite} %needed for reference ranges

\usepackage{enumitem}
\setlist{itemsep=0pt, topsep=1.5mm, leftmargin=8mm}

\newcommand{\pp}{\ensuremath{\mathrm {p\kern-0.05em p}}}
\newcommand{\PbPb}{\ensuremath{\mbox{Pb--Pb}}}
\newcommand{\GeVc}{\ensuremath{\mathrm{GeV}\kern-0.05em/\kern-0.02em c}}
\newcommand{\sqrts}{\ensuremath{\sqrt{s_{\mathrm{NN}}}}}
\newcommand{\pT}{\ensuremath{p_{\mathrm{T}}}}
\newcommand{\pTsub}{\ensuremath{p_{\mathrm{T,subleading}}}}
\newcommand{\pTlead}{\ensuremath{p_{\mathrm{T,leading}}}}
\newcommand{\kT}{\ensuremath{k_{\mathrm{T}}}}

\newcommand{\pTjet}{\ensuremath{p_{\mathrm{T,\;ch\; jet}}}}

\newcommand{\pTdet}{\ensuremath{p_{\mathrm{T,det}}^{\mathrm{ch\; jet}}}}

\newcommand{\tg}{\ensuremath{\theta_{\mathrm{g}}}}

\newcommand{\zg}{\ensuremath{z_{\mathrm{g}}}}

\newcommand{\zcut}{\ensuremath{z_{\mathrm{cut}}}}
\newcommand{\etajet}{\ensuremath{\eta_{\mathrm{jet}}}}
\newcommand{\Ninc}{\ensuremath{N_{\mathrm{jet, inc}}}}
\newcommand{\Sinc}{\ensuremath{\sigma_{\mathrm{jet, inc}}}}

\newcommand{\Rmax}{\ensuremath{R_{\mathrm{max}}}}
\newcommand{\Lres}{\ensuremath{L_{\mathrm{res}}}}

\begin{document}

%\linenumbers

%%%%%%%%%%%%%%%  Title page %%%%%%%%%%%%%%%%%%%%%%%%
\begin{titlepage}
\PHyear{2021}
\PHnumber{151}      % required, will be obtained from PH
\PHdate{23 July}  % required, will be obtained from PH
%

%%% Put your own title + short title here:
\title{Measurement of the groomed jet radius and momentum splitting fraction in pp and \PbPb{} collisions at $\bf{\sqrt{\textit{s}_{NN}}=5.02}$ TeV}
\ShortTitle{Measurement of the groomed jet radius and momentum splitting fraction}   % appears on right page headers

%%% Do not change the next lines
\Collaboration{ALICE Collaboration\thanks{See Appendix~\ref{app:collab} for the list of collaboration members}}
\ShortAuthor{ALICE Collaboration} % appears on left page headers, do not change

%\date{\today}
\begin{abstract}
This article presents groomed jet substructure measurements in \pp{} and \PbPb{} collisions
at $\sqrts=5.02$ TeV with the ALICE detector.
The Soft Drop grooming algorithm provides access to the hard parton splittings inside a jet by removing soft wide-angle radiation. We report the groomed jet momentum splitting fraction, 
$\zg$, and the (scaled) groomed jet radius, $\tg$.
Charged-particle jets are reconstructed at midrapidity using the anti-\kT{} algorithm 
with resolution parameters $R=0.2$ and $R=0.4$.
In heavy-ion collisions, the large underlying event poses
a challenge for the reconstruction of groomed jet observables, 
since fluctuations in the background can cause groomed parton splittings
to be misidentified.
By using strong grooming conditions to reduce this background,
we report these observables fully corrected for detector effects and background fluctuations for the first time. 
A narrowing of the \tg{} distribution in \PbPb{} 
collisions compared to \pp{} collisions is seen, 
which provides direct evidence of the modification of
the angular structure of jets
in the quark--gluon plasma. 
No significant modification of the $\zg$ distribution
in \PbPb{} collisions compared to \pp{} collisions is observed.
These results are compared with a variety of theoretical
models of jet quenching, and provide constraints on jet
energy-loss mechanisms and coherence effects in the quark--gluon plasma.
\end{abstract}

\end{titlepage}
\setcounter{page}{2}

%\maketitle

%%%%%%%%%%%%%%%%%%%%%%%%%%%%%%%%%%%%%%%%%%%%%%%%%%%%%%%%%%%%%%%%%%%%%%%%%%%%%%
%{\bf Introduction.} 
\section{Introduction}
Ultrarelativistic heavy-ion collisions at the Large Hadron Collider (LHC) are used to study the
high temperature deconfined phase of strongly interacting matter known as the quark--gluon plasma (QGP)~\cite{Adams:2005dq, Adcox:2004mh, LHC1review, Braun-Munzinger:2015hba, TheBigPicture}.
Highly-energetic jets created early in the collisions interact with the QGP medium and through those interactions they can lose energy and their internal structure can be modified.
This process,
known as jet quenching,
can be used to reveal the physical properties of the QGP itself such as its transport coefficients and the quasi-particle nature of its degrees of freedom as a function of scale \cite{PhysRevD.27.140, ReviewXinNian, ReviewYacine, ReviewMajumder}.
Experimentally, jet quenching is evaluated by comparing jet measurements in heavy-ion collisions to analogous measurements in
\pp{} collisions~\cite{PhysRevC.101.034911, atlas502, jetRaa276CMS, Adam:2020wen, hjetPbPb, hjetAuAu, AliceJetShape, atlasFF502, 2014243}. 
Notably, measurements of the jet angularity~\cite{AliceJetShape} and jet transverse profile~\cite{2014243}, which are sensitive
to a combination of the angular and momentum space structure of jets, suggest a narrowing of the jet core in heavy-ion collisions. Nonetheless, up to now, no direct modification of the intra-jet 
angular distribution alone has been measured.

Jet grooming algorithms provide access to the hard (high-momentum transfer) parton splittings inside a jet by removing soft wide-angle radiation~\cite{Larkoski:2014wba, Dasgupta:2013ihk, Larkoski:2015lea}.
Access to the hard splittings isolates substructures that are
well-controlled in perturbative QCD (pQCD), which in heavy-ion collisions
may help constrain various jet quenching effects such as
energy loss, transverse-momentum broadening, 
and color coherence.
Measurements of groomed jet observables in heavy-ion collisions have been performed by the ALICE and CMS collaborations ~\cite{PhysRevLett.120.142302, Acharya:2019djg, Sirunyan2018}, and opened a new avenue in 
the study of jet substructure in heavy-ion collisions.

\begin{figure}[!t]
\centering{}
\includegraphics[scale=0.39]{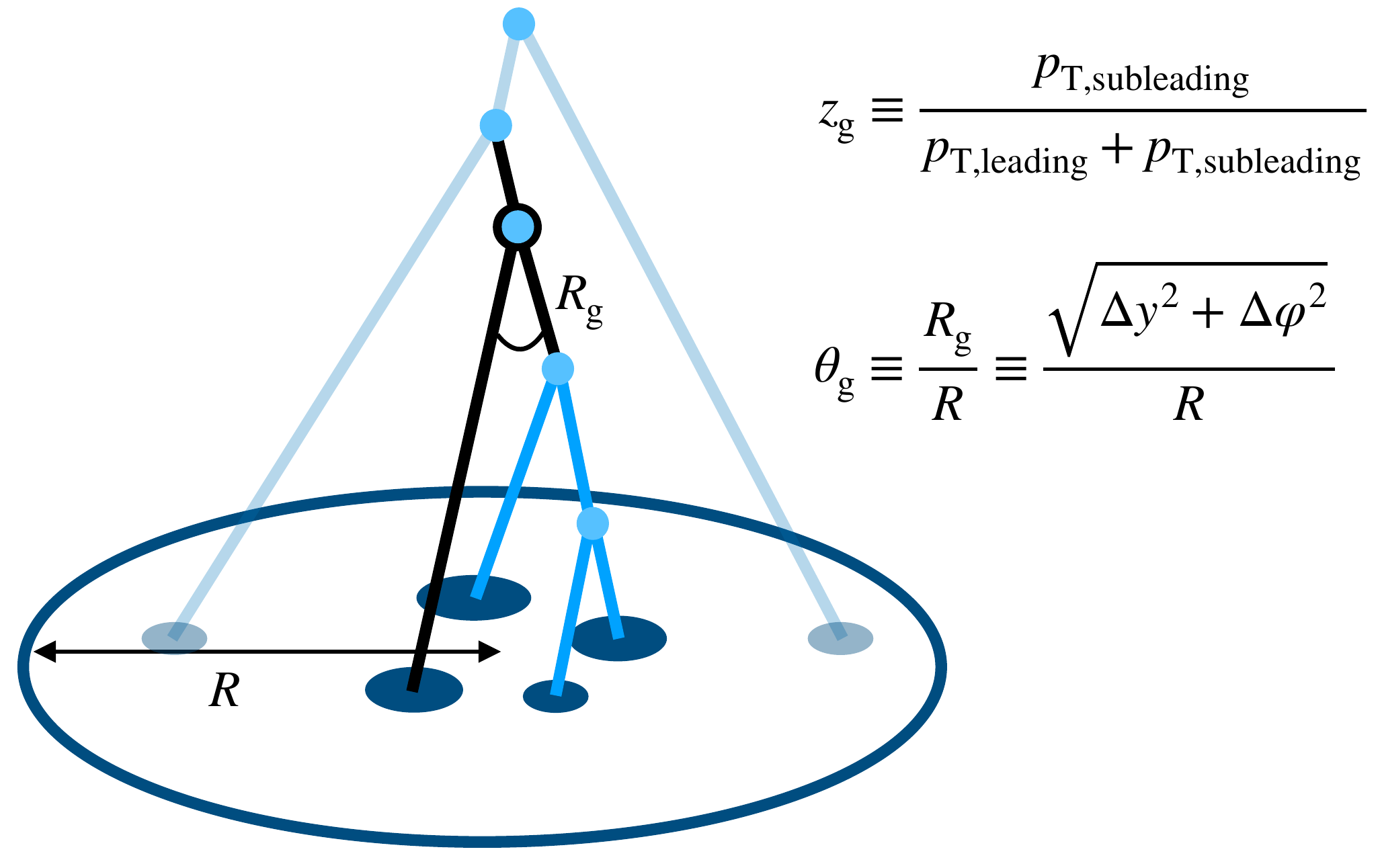}
\caption{Graphical representation of the angularly ordered Cambridge-Aachen reclustering of jet constituents
and subsequent Soft Drop grooming procedure~\cite{Larkoski:2014wba},
with the identified splitting denoted in black and the splittings that were groomed away in light blue.}
\label{fig:diagram}
\end{figure}

The Soft Drop (SD) \cite{Larkoski:2014wba, Dasgupta:2013ihk, Larkoski:2015lea} grooming algorithm identifies a single splitting by first reconstructing a jet with the anti-$k_{\rm T}$ algorithm and then reclustering the constituents of the jet using the Cambridge/Aachen (C/A) algorithm \cite{Dokshitzer:1997in} in order to follow the angular ordering of the QCD parton shower.
The splitting is selected from within the history of the reclustering with a grooming condition, $z > \zcut \theta^\beta$, where 
$\beta$ and \zcut\ are tunable parameters, 
$z$ is the fraction of transverse momentum (\pT) carried by the sub-leading (lowest \pT{}) prong,
\begin{equation} \label{eq:2}
z \equiv \frac{\pTsub} {\pTlead + \pTsub},
\end{equation}
and $\theta$ is the relative angular distance between the leading and sub-leading prong, 
\begin{equation} \label{eq:1}
\theta \equiv \frac{\sqrt{\Delta y ^2 + \Delta \varphi ^2}}{R},
\end{equation}
where $\Delta y$ and $\Delta \varphi$ are the distances measured in rapidity and azimuthal angle, respectively, and $R$ is the jet resolution parameter.
The groomed splitting is then characterized by two relevant kinematic observables: 
the groomed momentum fraction, \zg{}, and the (scaled) groomed jet radius, \tg{}, which are the values of $z$ and $\theta$  of Eq. (\ref{eq:2}) and (\ref{eq:1}) for the identified splitting, as shown in Fig.~\ref{fig:diagram}.

In \pp{} collisions, measurements of the \tg{} and \zg{} distributions were performed at RHIC and the LHC~\cite{PhysRevD.101.052007, PhysRevD.98.092014, Tripathee:2017ybi, Acharya:2019djg, STAR2020}.
At high-transverse momentum \pT{}, the data are described within uncertainties by pQCD
predictions~\cite{Kang:2019prh}.

In heavy-ion collisions, it was proposed that \tg{} may be sensitive to several important jet-quenching physics mechanisms: the relative suppression of gluon vs. quark jets, transverse-momentum broadening, and the ability of the medium to resolve a color dipole as two independent color charges
%, breaking color coherence 
\cite{Ringer:2019rfk, CasalderreySolana:2012ef}.
Uncertainty principle arguments suggest that
wider splittings are formed earlier in vacuum than narrower splittings 
($t_{\rm f} \sim 1/\tg^2$ where $t_{\rm f}$ is the splitting formation time). In heavy-ion collisions, this would result in
wider splittings traversing a longer path in the medium on average.
Complementary to \tg{}, it has been argued that \zg{} may be sensitive to 
the modification of the DGLAP splitting function in the QGP,
the breaking of color-coherence, and the response of the medium 
to the jet
\cite{Chien:2016led, Caucal:2019uvr, Chang:2019nrx, Casalderrey-Solana:2019ubu}.
By measuring both \tg{} and \zg{} simultaneously,
and thereby both the angular and momentum scales of the hard substructure of jets, these jet quenching mechanisms can be further constrained.

Up to now, no measurement of \tg{} has been performed in heavy-ion collisions.
Previous measurements of the \zg{} distribution by the CMS~\cite{PhysRevLett.120.142302} and ALICE collaborations~\cite{Acharya:2019djg} indicated significant modification with respect to \pp{} collisions. 
However, these results were not corrected for background and detector effects,
and are difficult to compare directly to theoretical
calculations
~\cite{mulligan2020identifying}. 
In this letter, we report the first fully corrected measurement of groomed substructure observables in heavy-ion collisions, allowing for a rigorous comparison with theoretical calculations.

%%%%%%%%%%%%%%%%%%%%%%%%%%%%%%%%%%%%%%%%%%%%%%%%%%%%%%%%%%%%%%%%%%%%%%%%%%%%%%
%{\bf Experimental setup and data sets.}
\section{Experimental setup and data sets}
A description of the ALICE detector and its performance can be found in
Refs.~\cite{aliceDetector, Abelev:2014ffa}. The \pp{} data set used in this
analysis was collected in 2017 during LHC Run 2 at $\sqrt{s} = 5.02$ TeV using a minimum-bias (MB) trigger defined by the coincidence of the signals from two
scintillator arrays in the forward region (V0 detectors)
~\cite{Abbas:2013taa}. The \PbPb{} data set was collected in 2018 at $\sqrts=5.02$ TeV. Central and semi-central triggers that select events in the 0--10\% and 30--50\%
centrality intervals based on the
multiplicity of produced particles in the forward V0 detectors, were used~\cite{centrality276, centrality502}.
The event selection includes a primary-vertex
selection and the removal of beam-induced background events and pileup~\cite{PhysRevC.101.034911}.
After these selections, the pp data sample
contains 870 million events and corresponds to an integrated luminosity of
$18.0 \pm 0.4$ nb$^{-1}$~\cite{ppXsec}. The \PbPb{} data sample contains 92 million events in
central collisions and 90 million events in semi-central collisions,
corresponding to an integrated luminosity of 0.12 nb$^{-1}$ and 0.06
nb$^{-1}$, respectively. 

This analysis uses charged-particle tracks reconstructed using information from both the Time Projection Chamber
(TPC)~\cite{Alme_2010} and the Inner Tracking System (ITS) \cite{Aamodt:2010aa}.
While track-based observables are collinear-unsafe~\cite{Chang:2013rca, Chen:2020vvp, Chien:2020hzh},
they can be measured with greater precision than calorimeter-based observables
and recent measurements have demonstrated that for the groomed jet observables considered here, track-based distributions are compatible with the corresponding collinear-safe distributions \cite{ATLAS:2019mgf}.
Tracks with $0.15 < \pT < 100 \;\GeVc$ were 
accepted over pseudorapidity range $|\eta| < 0.9$. Further details about the track selection are described in Ref.~\cite{THE_PUBLIC_NOTE}. 
The accepted tracks exhibit approximately uniform
azimuthal acceptance and momentum resolution $\sigma(\pT)/\pT$
ranging from about 1\% at track $\pT=1$ \GeVc{} to 4\% at track ${\pT=50 \;\GeVc}$.

%%%%%%%%%%%%%%%%%%%%%%%%%%%%%%%%%%%%%%%%%%%%%%%%%%%%%%%%%%%%%%%%%%%%%%%%%%%%%%
%{\bf Analysis method.}
\section{Analysis method}
%%%%%%%%%%%%%%%%%%%%%%%%%%%%%%%%%%%%%%%%%%%%%%%%%%%%
%{\it Jet reconstruction.}
Jets were reconstructed from charged-particle tracks with FastJet 3.2.1~\cite{Cacciari:2011ma} using the
anti-$k_{\mathrm{T}}$ algorithm with $E$-scheme recombination
for resolution parameters $R=0.2$ and $0.4$~\cite{antikt, catchment}. The pion mass is assumed for all jet constituents.
Jets in heavy-ion collisions have a large uncorrelated background contribution due
to fluctuations in the underlying event (UE)~\cite{Abelev2012}.
The event-by-event constituent subtraction method was used, which corrects the
overall jet \pT{} and its substructure simultaneously by subtracting
UE energy constituent by constituent~\cite{Berta:2014eza, Berta:2019hnj}. 
A maximum recombination distance $\Rmax = 0.25$ was used. After background
subtraction, the measured range is $40< \pTjet{} < 120$ GeV$/c$. 
The jet axis is required to be within the fiducial volume of the TPC, $\left| \etajet \right| < 0.9 - R$, where $\etajet$ is the jet pseudorapidity. 

Local background fluctuations in a heavy-ion collision environment can
result in 
an incorrect splitting (unrelated to the hard scattering) being 
identified by the grooming algorithm.
In order to address this issue, the measurement was
performed by applying a strong grooming condition,
$\zcut = 0.2$ (with $\beta =0$), which better mitigates
these effects as compared to softer grooming conditions (e.g. $\zcut=0.1$)~\cite{mulligan2020identifying}.
To further reduce the mistagging effects, we report measurements with either a small resolution parameter ($R=0.2$ in central collisions) or with more peripheral collisions (30--50\% for $R = 0.4$).

The rate of prong mistagging from residual background effects was evaluated by embedding 
jets simulated with the PYTHIA8 event generator~\cite{pythia} into measured \PbPb{} data and following the procedure in Ref.~\cite{mulligan2020identifying}.
The residual background
contribution ranges from approximately 5\% up to 15\% at lower \pT{}, in more central events, and at larger $R$. 
This level of background contamination is small enough to allow
the results to be unfolded for detector effects and background fluctuations.
The impact of the residual background contribution remains one of the main sources of systematic uncertainty~\cite{THE_PUBLIC_NOTE}.

%%%%%%%%%%%%%%%%%%%%%%%%%%%%%%%%%%%%%%%%%%%%%%%%%%%%
%{\it Unfolding.}
The reconstructed \pTjet{}, \tg{}, and \zg{} distributions were corrected for effects related to the tracking inefficiency,
particle-material interactions, and track \pT{} resolution. Moreover,
in \PbPb{} collisions, background fluctuations significantly smear
the reconstructed distributions of \tg{} and \zg{}. To account for
these effects, events were simulated with the PYTHIA8 generator using the Monash 2013 tune~\cite{pythia} and the GEANT3 model~\cite{Brun:1119728} for the particle transport in the ALICE detectors' material. For the \PbPb{} data, we embedded the simulated events into measured \PbPb{} data to mimic the background effects.
A four-dimensional response matrix describing
the detector and background response in \pTjet{} and \tg{} or \zg{} was constructed and used in the two-dimensional unfolding in \pTjet{}, \tg{} or \zg{} using the iterative Bayesian
unfolding algorithm~\cite{DAgostini, roounfold}.

%%%%%%%%%%%%%%%%%%%%%%%%%%%%%%%%%%%%%%%%%%%%%%%%%%%%%%%%%%%%%%%%%%%%%%%%%%%%%%
%{\bf Systematic uncertainties.}
\section{Systematic uncertainties}
The largest systematic uncertainties in this measurement originate from the tracking
inefficiency, the unfolding procedure, residual mistagged prongs, and the background subtraction procedure.
The total systematic uncertainty is calculated as the quadratic sum of all of the individual systematic uncertainties described below.

The systematic uncertainty due to the uncertainty of the tracking efficiency is evaluated using random rejection of additional tracks in jet finding according to the estimated tracking efficiency uncertainty of 4\%,
based on variations in the track selection criteria and on the ITS-TPC
track-matching efficiency uncertainty. 
The systematic uncertainty arising from the unfolding regularization procedure is evaluated by
varying the number of unfolding iterations by $\pm2$ units, 
scaling the prior distribution,
varying the binning, and varying the lower bound in the detector-level charged-particle jet transverse momentum \pTdet{} range by 5 \GeVc.
The systematic uncertainty due to the model-dependence of the generator used to construct the response matrix is estimated by comparing results obtained with PYTHIA~\cite{pythia}, HERWIG~\cite{Bellm:2015jjp}, and JEWEL \cite{Zapp:2013vla}.
The systematic uncertainty due to the bias introduced by the constituent subtraction
procedure is estimated by varying \Rmax{} from ``under-subtraction'' ($\Rmax=0.05$)
to ``over-subtraction'' ($\Rmax=0.7$), around the nominal value of $\Rmax=0.25$. 
The systematic uncertainty due to a possible residual contamination of mistagged splittings
after unfolding is estimated with a closure test. The total relative systematic uncertainty ranges from 3--24\% for \tg{} and 4--10\% for \zg{}. See Ref.~\cite{THE_PUBLIC_NOTE} for more details about the systematic uncertainties used in this measurement.
%%%%%%%%%%%%%%%%%%%%%%%%%%%%%%%%%%%%%%%%%%%%%%%%%%%%%%%%%%%%%%%%%%%%%%%%%%

\begin{figure*}[!ht]
\centering{}
\includegraphics[width=0.48\textwidth]{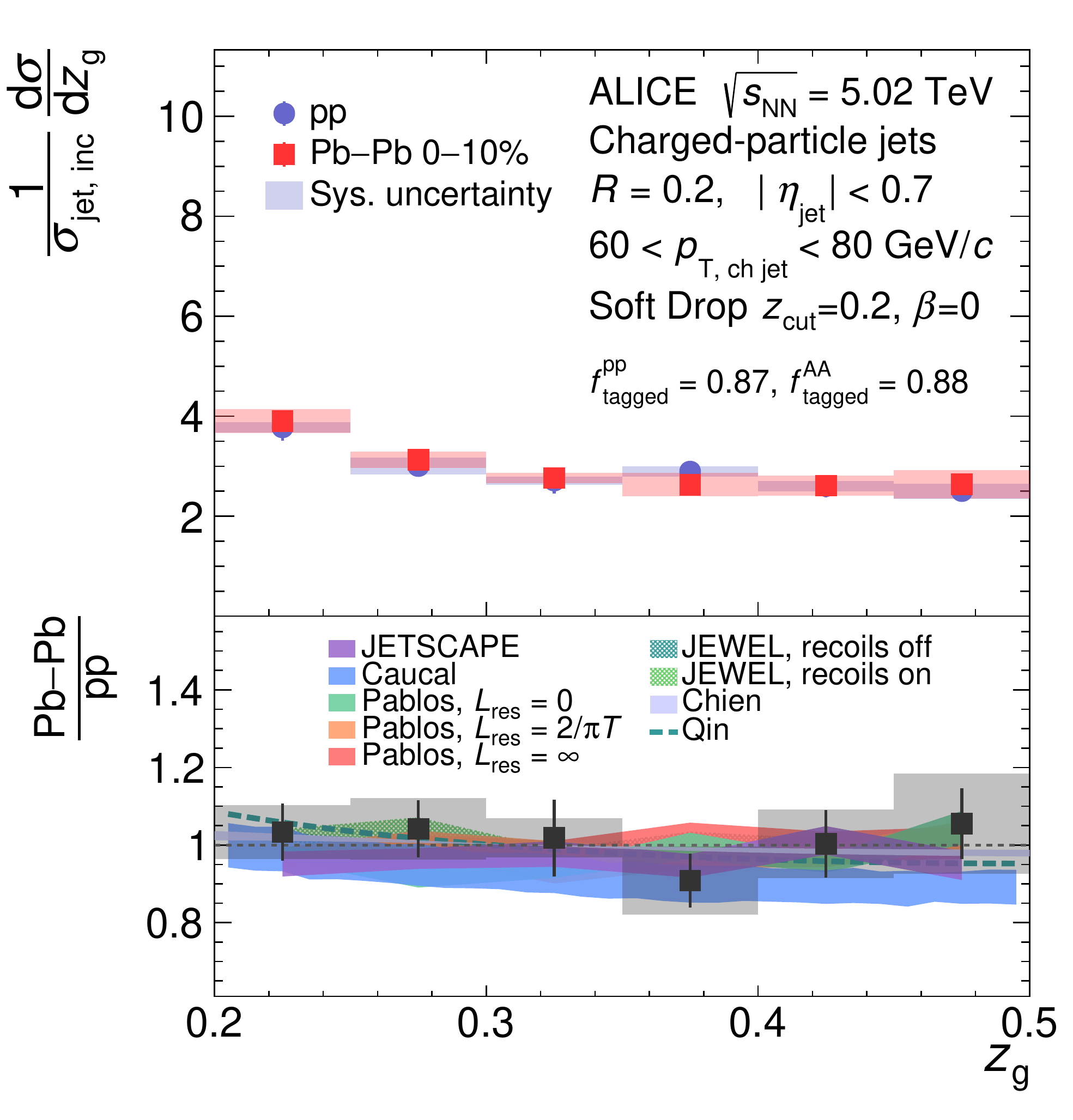}
\includegraphics[width=0.48\textwidth]{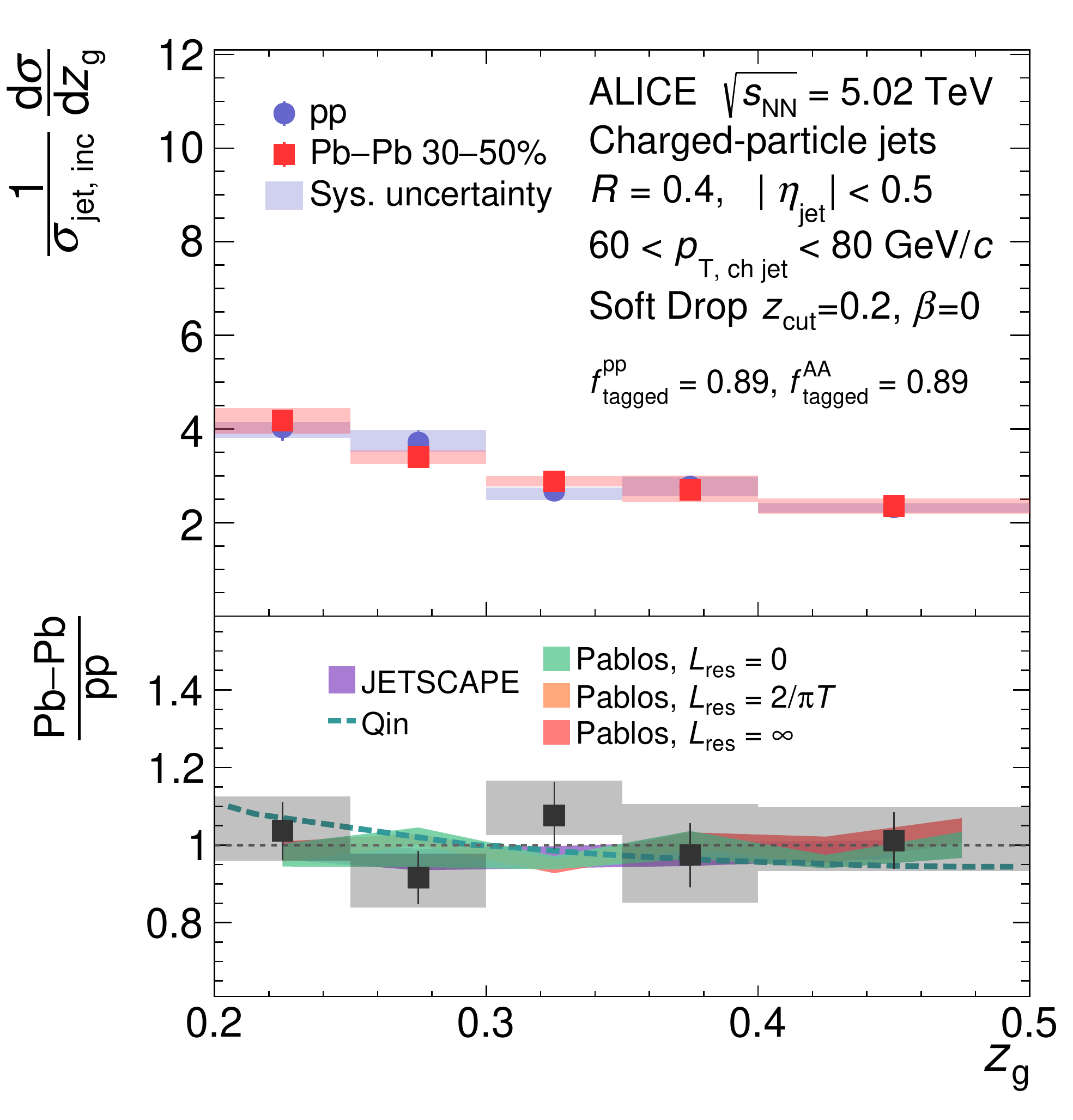}
\caption{Unfolded \zg{} distributions for charged-particle jets in \pp{} collisions compared to those in \PbPb{} collisions at $\sqrts=5.02$ TeV with $\zcut = 0.2$
for 0--10\% centrality for $R=0.2$ (left)
 and  30--50\% centrality for $R=0.4$ (right). 
 The distributions are normalized to the inclusive jet cross section in the $60 < \pTjet < 80$ GeV/$c$ interval, and
 $f_{\mathrm{tagged}}$ indicates the fraction of splittings that were tagged to pass the SD condition in the selected \pTjet{} interval. The ratios in the bottom panel are compared to the following theoretical predictions: JETSCAPE~\cite{Putschke:2019yrg}, JEWEL~\cite{Zapp:2012ak, Zapp:2013vla}, Caucal et al.~\cite{Caucal:2019uvr, Caucal_2018}, Chien et al.~\cite{Chien:2016led}, Qin et al.~\cite{Chang:2019nrx}, and Pablos et al.~\cite{HybridModel, HybridModelResolution, Casalderrey-Solana:2019ubu}. Further details can be found in Ref. \cite{THE_PUBLIC_NOTE}.}
\label{fig:zg}
\end{figure*}

%%%%%%%%%%%%%%%%%%%%%%%%%%%%%%%%%%%%%%%%%%%%%%%%%%%%%%%%%%%%%%%%%%%%%%%%%%%%%%
%{\bf Results.}
\section{Results}
We report the \tg{} and \zg{} distributions in
the \pTjet{} interval between 60 and 80 GeV/$c$ for $\zcut{} = 0.2$ in
central (0--10\%, $R=0.2$) and  semi-central (30--50\%, $R=0.4$) \PbPb{} collisions.
The distributions are reported as normalized differential cross sections,
\begin{equation} \label{eq:7}
\frac{1}{\Sinc} \frac{\rm{d}\sigma}{\rm{d}\zg}
=
\frac{1}{\Ninc} \frac{\mathrm{d}N} {\rm{d}\zg},
\end{equation}
where $N$ is the number of jets passing the SD 
condition with a given \pTjet{}, \Ninc{} is the number of inclusive jets, 
and $\sigma,\Sinc$ are the corresponding cross sections.
The analog of Eq. (\ref{eq:7}) also applies for \tg{}.

\begin{figure*}[!ht]
\centering{}
\includegraphics[width=0.48\textwidth]{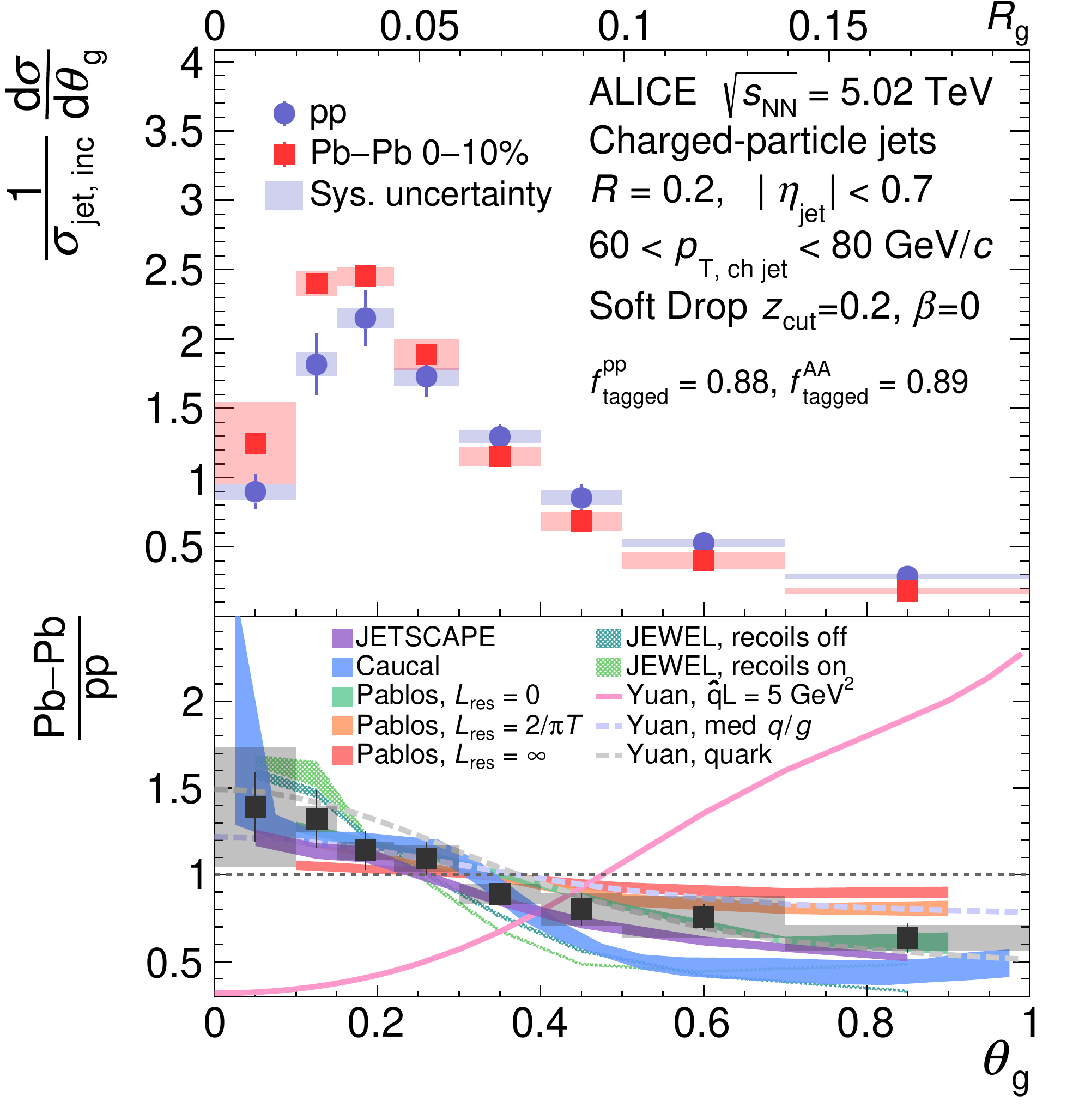}
\includegraphics[width=0.48\textwidth]{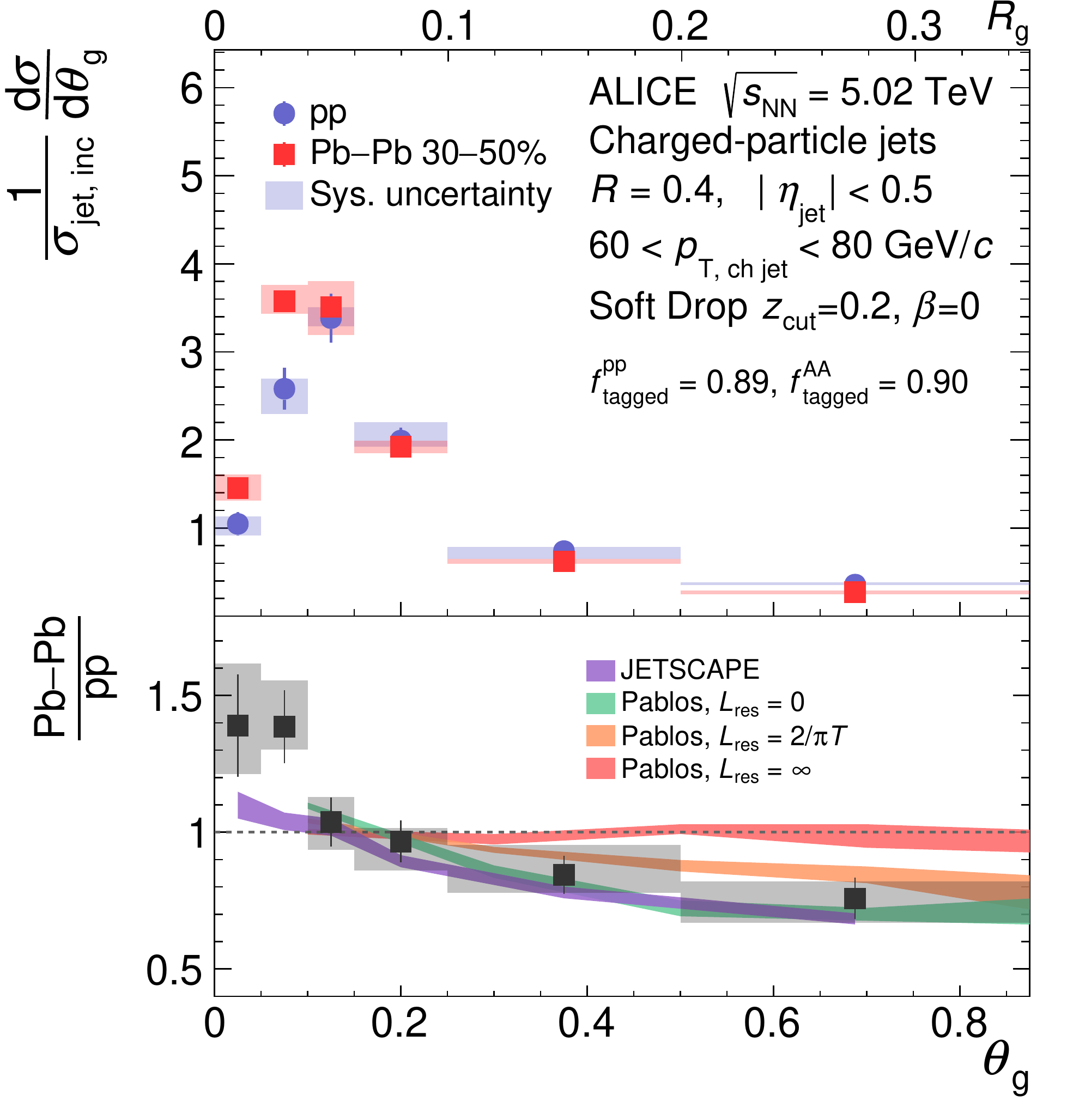}
\caption{Unfolded \tg{} distributions for charged-particle jets in \pp{} 
collisions compared to those in \PbPb{} collisions at $\sqrts=5.02$ TeV with $\zcut = 0.2$ for 
0--10\% centrality for $R=0.2$ (left) and  
30--50\% centrality for $R=0.4$ (right). 
The distributions are normalized to the inclusive jet cross section in the $60 < \pTjet < 80$ GeV/$c$ interval, and
 $f_{\mathrm{tagged}}$ indicates the fraction of splittings that were tagged to pass the SD condition in the selected \pTjet{} interval.
 The ratios in the bottom panel are compared to the following theoretical predictions: JETSCAPE~\cite{Putschke:2019yrg}, JEWEL~\cite{Zapp:2012ak, Zapp:2013vla}, Caucal et al.~\cite{Caucal:2019uvr, Caucal_2018}, Pablos et al.~\cite{HybridModel, HybridModelResolution, Casalderrey-Solana:2019ubu}, and Yuan et al.~\cite{Ringer:2019rfk}. Further details can be found in Ref. \cite{THE_PUBLIC_NOTE}.}
\label{fig:tg}
\end{figure*}

The \zg{} and \tg{} distributions are shown in Fig.~\ref{fig:zg}
and Fig.~\ref{fig:tg}, respectively.
The distributions from \PbPb{} collisions are compared with the corresponding distributions from pp collisions,
with their ratios displayed in the bottom panels.
The relative uncertainties are assumed to be uncorrelated between \pp{} and \PbPb{} collisions, and are added in quadrature in the ratio.
In \PbPb{} collisions the precision of the measurements decreases as the jet
resolution parameter is increased or the centrality is decreased, as the prong mistagging probability decreases with centrality and with decreasing $R$.

The fraction of jets that do not contain a splitting which
passes the SD condition ($f_{\mathrm{tagged}}$)
differs by at most 1\% between \PbPb{} and \pp{} collisions.
Therefore, any modifications in \PbPb{} compared to \pp{} collisions 
can change the shape of the distribution, but keep the integral approximately the same.

The \zg{} distributions in \PbPb{} and \pp{} collisions are consistent within experimental uncertainties for all jet momenta, jet resolution parameters, and centralities measured.

The situation is remarkably different when comparing the groomed jet radius, \tg{}, in both systems.
For $R=0.2$ in central collisions, the data suggests a narrowing of the \PbPb{} distribution relative to the \pp{} distribution is observed.
This narrowing persists even in semi-central \PbPb{} collisions for $R=0.4$ 
where quenching effects are expected to be less than in central collisions.

%%%%%%%%%%%%%%%%%%%%%%%%%%%%%%%%%%%%%%%%%%%%%%%%%%%%%%%%%%%%%%%%%%%

%%%%%%%%%%%%%%%%%%%%%%%%%%%%%%%%%%%%%%%%%%%%%%%%%%%%%%%%%%%%%
%{\it Comparing to theoretical models.}
We compare the ratio of the measurements in pp and \PbPb{} collisions with several theoretical implementations
of jet quenching:

\begin{itemize}

\item {\it JETSCAPE} 
\cite{Putschke:2019yrg} consists of a medium-modified parton
shower with the MATTER model~\cite{Majumder_2013} controlling the high-virtuality 
phase and the Linear Boltzmann Transport (LBT) model describing the low-virtuality phase~\cite{LBT}. The version of JETSCAPE used for this calculation employs a jet transport coefficient, $\hat{q}$, that includes dependence on parton virtuality, in addition to dependence on the local temperature and running of the parton-medium coupling.

\item {\it JEWEL}
\cite{Zapp:2012ak, Zapp:2013vla} consists a Monte Carlo implementation of
BDMPS-based medium-induced gluon radiation in a medium
modeled with a Bjorken expansion.
We consider the impact of medium recoil by including calculations both with and without recoils enabled~\cite{KunnawalkamElayavalli:2017hxo}.

\item {\it Caucal et al.}~\cite{Caucal:2019uvr, Caucal_2018} implements a pQCD parton
shower with incoherent interactions including both factorized vacuum and medium-induced emissions
in a static brick medium. 

\item {\it Chien et al.}~\cite{Chien:2016led} (for \zg{} only)
applies Soft Collinear Effective Theory with Glauber gluon interactions. 

\item {\it Qin et al.}~\cite{Chang:2019nrx} 
(for \zg{} only) applies the higher twist formalism with coherent energy loss. 

\item {\it Pablos et al.} 
\cite{HybridModel, HybridModelResolution, Casalderrey-Solana:2019ubu}
consists of partons produced by a vacuum shower that interact
with the medium according to a strongly-coupled
AdS/CFT-based model.
The parameter \Lres{} describes the
degree to which the medium can resolve the jet angular
structure, where $\Lres = 0$ corresponds to full
resolution of all jet constituents (fully incoherent),
$\Lres = \infty$ corresponds to fully coherent energy loss,
and $\Lres=2/\pi T$ is an intermediate case, where $T$ is the local medium temperature. 

\item {\it Yuan et al.}~\cite{Ringer:2019rfk} 
(for \tg{} only)
``med $q$/$g$'' and ``quark'' consist of 
medium-modified quark-gluon fractions 
without any additional effects, 
where the quark-gluon fractions in the ``med $q$/$g$'' case are 
extracted in Ref.~\cite{Qiu:2019sfj} with a relative suppression
factor of approximately four between gluon jets and quark jets.
The calculation labeled 
``$\hat{q}L$'' 
includes an implementation of transverse-momentum broadening. 

\end{itemize}

The \PbPb-to-pp{} ratios of the \zg{} distributions are consistent with all theoretical
predictions considered. The predicted modifications,
which have been constrained by previous measurements
\cite{PhysRevLett.120.142302, Acharya:2019djg}, 
are small, and the differences between them are
yet smaller than the current uncertainty of the data.
Nevertheless, these new measurements are the first direct
comparisons of predictions to fully corrected data, 
and limit the possible 
in-medium modifications of the momentum structure of hard splittings to be less 
than 10--20\% depending on the centrality, jet $R$, and the grooming
settings considered.

Despite employing different microscopic implementations of the jet-medium interactions, the majority of the models capture the qualitative feature of the narrowing seen in the 
\tg{} distributions.
The theoretical models can be grouped according to three distinct
mechanisms by which \tg{} is modified: incoherent energy loss, 
coherent energy loss, and transverse broadening. 
The measurements 
are consistent with models implementing (transverse) incoherent interaction of the jet shower constituents with the medium. This is illustrated by calculations of Pablos et al. where the data favor the incoherent energy loss ($\Lres=0$) and is also supported by Caucal et al., JEWEL, and JETSCAPE.
On the other hand, the Yuan et al. calculation with
medium-modified ``quark--gluon'' fractions indicates that the data could be explained by the stronger suppression of gluon showers, which are on average broader, with fully coherent energy loss.
These two physics mechanisms --- the degree of 
incoherent energy loss, and the relative quark/gluon 
suppression --- both lead to a 
suppression of wide-angle splittings.
The prediction by Yuan et al. ``$\hat{q}L$'' exhibits the opposite
trend compared to the data, demonstrating that there is no
strong transverse broadening in the hard substructure.

The presented measurements indicate that the medium has a significant resolving power for splittings with a particular dependence on the angular (or coherence) scale, promoting narrow structures or filtering out 
wider jets altogether.

%%%%%%%%%%%%%%%%%%%%%%%%%%%%%%%%%%%%%%%%%%%%%%%%%%%%%%%%%%%%%
%{\bf Conclusions.}
\section{Conclusions}
We reported the groomed jet momentum fraction, 
$\zg$, and the (scaled) groomed jet radius, $\tg$, 
of charged-particle jets measured in \pp{} and \PbPb{} collisions
at $\sqrts=5.02$ TeV with the ALICE detector.
By using stronger grooming conditions in the
SD grooming algorithm, we suppressed contamination
of mistagged splittings from the underlying event, 
and unfolded the final distributions 
for detector and background fluctuation effects.
This allows for the first time the direct comparison of groomed jet measurements 
in heavy-ion collisions with theoretical predictions 
of jet quenching in the QGP.
The \zg{} distributions are consistent with no modification
in \PbPb{} collisions compared to \pp{} collisions.
The \tg{} distributions are narrower in 
\PbPb{} collisions compared to \pp{} collisions,
which is the first direct experimental evidence for the modification of
the angular scale of groomed jets in heavy-ion collisions.

These new results demonstrate sensitivity to the microscopic structure of the QGP, including its angular resolving power.
This marks an important step towards quantitative understanding of the properties of the QGP, and provides a new path for novel differential
jet substructure measurements to further elucidate the microscopic nature of the QGP. 

%%%%%%%%%%%%%%%%%%%%%%%%%%%%%%%%
% end main text 
%%%%%%%%%%%%%%%%%%%%%%%%%%%%%%%%

%%%%% acknowledgements - handled by EB chairs 
\newenvironment{acknowledgement}{\relax}{\relax}
\begin{acknowledgement}
\section*{Acknowledgements}
% add specific acknowledgements here 
% ...but please don't remove the line below: funding agencies
% will be acknowledged with a custom tex file handled by EB chairs after Collab Round 2
We gratefully acknowledge Paul Caucal, Yang-Ting Chien, 
Daniel Pablos, Chanwook Park, Guang-You Qin, Gregory Soyez, 
Feng Yuan, and the JETSCAPE Collaboration
for providing theoretical predictions.
We thank Korinna Zapp for discussions regarding recoil subtraction in the JEWEL model.
% Version: 2021-07-15

The ALICE Collaboration would like to thank all its engineers and technicians for their invaluable contributions to the construction of the experiment and the CERN accelerator teams for the outstanding performance of the LHC complex.
The ALICE Collaboration gratefully acknowledges the resources and support provided by all Grid centres and the Worldwide LHC Computing Grid (WLCG) collaboration.
The ALICE Collaboration acknowledges the following funding agencies for their support in building and running the ALICE detector:
A. I. Alikhanyan National Science Laboratory (Yerevan Physics Institute) Foundation (ANSL), State Committee of Science and World Federation of Scientists (WFS), Armenia;
Austrian Academy of Sciences, Austrian Science Fund (FWF): [M 2467-N36] and Nationalstiftung f\"{u}r Forschung, Technologie und Entwicklung, Austria;
Ministry of Communications and High Technologies, National Nuclear Research Center, Azerbaijan;
Conselho Nacional de Desenvolvimento Cient\'{\i}fico e Tecnol\'{o}gico (CNPq), Financiadora de Estudos e Projetos (Finep), Funda\c{c}\~{a}o de Amparo \`{a} Pesquisa do Estado de S\~{a}o Paulo (FAPESP) and Universidade Federal do Rio Grande do Sul (UFRGS), Brazil;
Ministry of Education of China (MOEC) , Ministry of Science \& Technology of China (MSTC) and National Natural Science Foundation of China (NSFC), China;
Ministry of Science and Education and Croatian Science Foundation, Croatia;
Centro de Aplicaciones Tecnol\'{o}gicas y Desarrollo Nuclear (CEADEN), Cubaenerg\'{\i}a, Cuba;
Ministry of Education, Youth and Sports of the Czech Republic, Czech Republic;
The Danish Council for Independent Research | Natural Sciences, the VILLUM FONDEN and Danish National Research Foundation (DNRF), Denmark;
Helsinki Institute of Physics (HIP), Finland;
Commissariat \`{a} l'Energie Atomique (CEA) and Institut National de Physique Nucl\'{e}aire et de Physique des Particules (IN2P3) and Centre National de la Recherche Scientifique (CNRS), France;
Bundesministerium f\"{u}r Bildung und Forschung (BMBF) and GSI Helmholtzzentrum f\"{u}r Schwerionenforschung GmbH, Germany;
General Secretariat for Research and Technology, Ministry of Education, Research and Religions, Greece;
National Research, Development and Innovation Office, Hungary;
Department of Atomic Energy Government of India (DAE), Department of Science and Technology, Government of India (DST), University Grants Commission, Government of India (UGC) and Council of Scientific and Industrial Research (CSIR), India;
Indonesian Institute of Science, Indonesia;
Istituto Nazionale di Fisica Nucleare (INFN), Italy;
Institute for Innovative Science and Technology , Nagasaki Institute of Applied Science (IIST), Japanese Ministry of Education, Culture, Sports, Science and Technology (MEXT) and Japan Society for the Promotion of Science (JSPS) KAKENHI, Japan;
Consejo Nacional de Ciencia (CONACYT) y Tecnolog\'{i}a, through Fondo de Cooperaci\'{o}n Internacional en Ciencia y Tecnolog\'{i}a (FONCICYT) and Direcci\'{o}n General de Asuntos del Personal Academico (DGAPA), Mexico;
Nederlandse Organisatie voor Wetenschappelijk Onderzoek (NWO), Netherlands;
The Research Council of Norway, Norway;
Commission on Science and Technology for Sustainable Development in the South (COMSATS), Pakistan;
Pontificia Universidad Cat\'{o}lica del Per\'{u}, Peru;
Ministry of Education and Science, National Science Centre and WUT ID-UB, Poland;
Korea Institute of Science and Technology Information and National Research Foundation of Korea (NRF), Republic of Korea;
Ministry of Education and Scientific Research, Institute of Atomic Physics and Ministry of Research and Innovation and Institute of Atomic Physics, Romania;
Joint Institute for Nuclear Research (JINR), Ministry of Education and Science of the Russian Federation, National Research Centre Kurchatov Institute, Russian Science Foundation and Russian Foundation for Basic Research, Russia;
Ministry of Education, Science, Research and Sport of the Slovak Republic, Slovakia;
National Research Foundation of South Africa, South Africa;
Swedish Research Council (VR) and Knut \& Alice Wallenberg Foundation (KAW), Sweden;
European Organization for Nuclear Research, Switzerland;
Suranaree University of Technology (SUT), National Science and Technology Development Agency (NSDTA) and Office of the Higher Education Commission under NRU project of Thailand, Thailand;
Turkish Energy, Nuclear and Mineral Research Agency (TENMAK), Turkey;
National Academy of  Sciences of Ukraine, Ukraine;
Science and Technology Facilities Council (STFC), United Kingdom;
National Science Foundation of the United States of America (NSF) and United States Department of Energy, Office of Nuclear Physics (DOE NP), United States of America.
\end{acknowledgement}

%%%%%%%%
% References:
\bibliographystyle{utphys}
\bibliography{main.bib}

\providecommand{\href}[2]{#2}\begingroup\raggedright\begin{thebibliography}{10}

\bibitem{Adams:2005dq}
{\bfseries STAR} Collaboration, J.~Adams {\em et~al.}, ``{Experimental and
  theoretical challenges in the search for the quark gluon plasma: The STAR
  Collaboration's critical assessment of the evidence from RHIC collisions}'',
  \href{http://dx.doi.org/10.1016/j.nuclphysa.2005.03.085}{{\em Nucl. Phys. A}
  {\bfseries 757} (2005) 102--183},
  \href{http://arxiv.org/abs/nucl-ex/0501009}{{\ttfamily
  arXiv:nucl-ex/0501009}}.

\bibitem{Adcox:2004mh}
{\bfseries PHENIX} Collaboration, K.~Adcox {\em et~al.}, ``{Formation of dense
  partonic matter in relativistic nucleus-nucleus collisions at RHIC:
  Experimental evaluation by the PHENIX collaboration}'',
  \href{http://dx.doi.org/10.1016/j.nuclphysa.2005.03.086}{{\em Nucl. Phys. A}
  {\bfseries 757} (2005) 184--283},
  \href{http://arxiv.org/abs/nucl-ex/0410003}{{\ttfamily
  arXiv:nucl-ex/0410003}}.

\bibitem{LHC1review}
B.~M{\"u}ller, J.~Schukraft, and B.~Wys{\l}ouch, ``{First Results from Pb+Pb
  Collisions at the LHC}'',
  \href{http://dx.doi.org/10.1146/annurev-nucl-102711-094910}{{\em Annu. Rev.
  Nucl. Part. S.} {\bfseries 62} (2012) 361--386}.

\bibitem{Braun-Munzinger:2015hba}
P.~Braun-Munzinger, V.~Koch, T.~Sch{\"a}fer, and J.~Stachel, ``{Properties of
  hot and dense matter from relativistic heavy ion collisions}'',
  \href{http://dx.doi.org/10.1016/j.physrep.2015.12.003}{{\em Phys. Rept.}
  {\bfseries 621} (2016) 76--126}.

\bibitem{TheBigPicture}
W.~Busza, K.~Rajagopal, and W.~van~der Schee, ``{Heavy Ion Collisions: The Big
  Picture, and the Big Questions}'',
  \href{http://dx.doi.org/10.1146/annurev-nucl-101917-020852}{{\em Ann. Rev.
  Nucl. Part. Sci.} {\bfseries 68} (2018) 339--376}.

\bibitem{PhysRevD.27.140}
J.~D. Bjorken, ``Highly relativistic nucleus-nucleus collisions: The central
  rapidity region'', \href{http://dx.doi.org/10.1103/PhysRevD.27.140}{{\em
  Phys. Rev. D} {\bfseries 27} (1983) 140--151}.

\bibitem{ReviewXinNian}
G.-Y. Qin and X.-N. Wang, ``{Jet quenching in high-energy heavy-ion
  collisions}'', \href{http://dx.doi.org/10.1142/S0218301315300143}{{\em Int.
  J. Mod. Phys. E} {\bfseries 24} (2015) 1530014}.

\bibitem{ReviewYacine}
J.-P. Blaizot and Y.~Mehtar-Tani, ``{Jet structure in heavy ion collisions}'',
  \href{http://dx.doi.org/10.1142/S021830131530012X}{{\em Int. J. Mod. Phys. E}
  {\bfseries 24} (2015) 1530012}.

\bibitem{ReviewMajumder}
A.~Majumder and M.~van Leeuwen, ``{The theory and phenomenology of perturbative
  QCD based jet quenching}'',
  \href{http://dx.doi.org/https://doi.org/10.1016/j.ppnp.2010.09.001}{{\em
  Prog. Part. Nucl. Phys.} {\bfseries 66} (2011) 41--92}.

\bibitem{PhysRevC.101.034911}
{\bfseries ALICE} Collaboration, ``{Measurements of inclusive jet spectra in
  \pp{} and central \PbPb{} collisions at
  $\sqrts=5.02\phantom{\rule{0.16em}{0ex}}\mathrm{TeV}$}'',
  \href{http://dx.doi.org/10.1103/PhysRevC.101.034911}{{\em Phys. Rev. C}
  {\bfseries 101} (2020) 034911}.

\bibitem{atlas502}
{\bfseries ATLAS} Collaboration, ``{Measurement of the nuclear modification
  factor for inclusive jets in Pb+Pb collisions at $\sqrt{s_\mathrm{NN}}=5.02$
  TeV with the ATLAS detector}'',
  \href{http://dx.doi.org/10.1016/j.physletb.2018.10.076}{{\em Phys. Lett. B}
  {\bfseries 790} (2019) 108--128}.

\bibitem{jetRaa276CMS}
{\bfseries CMS} Collaboration, ``{Measurement of inclusive jet cross sections
  in \pp{} and \PbPb{} collisions at $\sqrts=2.76$ TeV}'',
  \href{http://dx.doi.org/10.1103/PhysRevC.96.015202}{{\em Phys. Rev. C}
  {\bfseries 96} (2017) 015202}.

\bibitem{Adam:2020wen}
{\bfseries STAR} Collaboration, J.~Adam {\em et~al.}, ``{Measurement of
  inclusive charged-particle jet production in Au + Au collisions at
  $\sqrt{s_{NN}}=$200 GeV}'',
  \href{http://dx.doi.org/10.1103/PhysRevC.102.054913}{{\em Phys. Rev. C}
  {\bfseries 102} no.~5, (2020) 054913},
  \href{http://arxiv.org/abs/2006.00582}{{\ttfamily arXiv:2006.00582
  [nucl-ex]}}.

\bibitem{hjetPbPb}
{\bfseries ALICE} Collaboration, ``{Measurement of jet quenching with
  semi-inclusive hadron-jet distributions in central Pb--Pb collisions at
  $\sqrt{{s}_{\rm NN}} = 2.76$ TeV}'',
  \href{http://dx.doi.org/10.1007/JHEP09(2015)170}{{\em J. High Energ. Phys.}
  {\bfseries 2015} (2015) 170}.

\bibitem{hjetAuAu}
{\bfseries STAR} Collaboration, ``{Measurements of jet quenching with
  semi-inclusive hadron+jet distributions in $\text{Au}+\text{Au}$ collisions
  at $\sqrt{{s}_{NN}}=200$ GeV}'',
  \href{http://dx.doi.org/10.1103/PhysRevC.96.024905}{{\em Phys. Rev. C}
  {\bfseries 96} (2017) 024905}.

\bibitem{AliceJetShape}
{\bfseries ALICE} Collaboration, ``{Medium modification of the shape of
  small-radius jets in central \PbPb{} collisions at $\sqrts=2.76$ TeV}'',
  \href{http://dx.doi.org/10.1007/JHEP10(2018)139}{{\em J. High Energ. Phys.}
  {\bfseries 2018} (2018) 139}.

\bibitem{atlasFF502}
{\bfseries ATLAS} Collaboration, ``{Measurement of jet fragmentation in \PbPb{}
  and \pp{} collisions at $\sqrts=5.02$ TeV with the ATLAS detector}'',
  \href{http://dx.doi.org/10.1103/PhysRevC.98.024908}{{\em Phys. Rev. C}
  {\bfseries 98} (2018) 024908}.

\bibitem{2014243}
{\bfseries CMS} Collaboration, ``{Modification of jet shapes in \PbPb{}
  collisions at $\sqrts=2.76$ TeV}'',
  \href{http://dx.doi.org/https://doi.org/10.1016/j.physletb.2014.01.042}{{\em
  Physics Letters B} {\bfseries 730} (2014) 243--263}.

\bibitem{Larkoski:2014wba}
A.~J. Larkoski, S.~Marzani, G.~Soyez, and J.~Thaler, ``{Soft Drop}'',
\href{http://dx.doi.org/10.1007/JHEP05(2014)146}{{\em J. High Energ. Phys.}
  {\bfseries 05} (2014) 146}.
%%CITATION = ARXIV:1402.2657;%%.

\bibitem{Dasgupta:2013ihk}
M.~Dasgupta, A.~Fregoso, S.~Marzani, and G.~P. Salam, ``{Towards an
  understanding of jet substructure}'',
\href{http://dx.doi.org/10.1007/JHEP09(2013)029}{{\em J. High Energ. Phys.}
  {\bfseries 09} (2013) 029}.
%%CITATION = ARXIV:1307.0007;%%.

\bibitem{Larkoski:2015lea}
A.~J. Larkoski, S.~Marzani, and J.~Thaler, ``{Sudakov Safety in Perturbative
  QCD}'',
\href{http://dx.doi.org/10.1103/PhysRevD.91.111501}{{\em Phys. Rev.} {\bfseries
  D91} (2015) 111501}.
%%CITATION = ARXIV:1502.01719;%%.

\bibitem{PhysRevLett.120.142302}
{\bfseries CMS} Collaboration, ``{Measurement of the Splitting Function in
  \pp{} and \PbPb{} Collisions at $\sqrts=5.02$ TeV}'',
  \href{http://dx.doi.org/10.1103/PhysRevLett.120.142302}{{\em Phys. Rev.
  Lett.} {\bfseries 120} (2018) 142302}.

\bibitem{Acharya:2019djg}
{\bfseries ALICE} Collaboration, ``{Exploration of jet substructure using
  iterative declustering in pp and Pb--Pb collisions at LHC energies}'',
  \href{http://dx.doi.org/10.1016/j.physletb.2020.135227}{{\em Phys. Lett. B}
  {\bfseries 802} (2020) 135227}.

\bibitem{Sirunyan2018}
{\bfseries CMS} Collaboration, ``{Measurement of the groomed jet mass in
  \PbPb{} and \pp{} collisions at $\sqrts=5.02$ TeV}'',
  \href{http://dx.doi.org/10.1007/JHEP10(2018)161}{{\em J. High Energ. Phys.}
  {\bfseries 2018} (2018) 161}.

\bibitem{Dokshitzer:1997in}
Y.~L. Dokshitzer, G.~D. Leder, S.~Moretti, and B.~R. Webber, ``{Better jet
  clustering algorithms}'',
  \href{http://dx.doi.org/10.1088/1126-6708/1997/08/001}{{\em JHEP} {\bfseries
  08} (1997) 001}, \href{http://arxiv.org/abs/hep-ph/9707323}{{\ttfamily
  arXiv:hep-ph/9707323}}.

\bibitem{PhysRevD.101.052007}
{\bfseries ATLAS} Collaboration, ``{Measurement of soft-drop jet observables in
  \pp{} collisions with the ATLAS detector at $\sqrts=13$ TeV}'',
  \href{http://dx.doi.org/10.1103/PhysRevD.101.052007}{{\em Phys. Rev. D}
  {\bfseries 101} (2020) 052007}.

\bibitem{PhysRevD.98.092014}
{\bfseries CMS} Collaboration, ``Measurement of jet substructure observables in
  $t\overline{t}$ events from proton-proton collisions at $\sqrt{s}=13\text{
  }\text{ }\mathrm{TeV}$'',
  \href{http://dx.doi.org/10.1103/PhysRevD.98.092014}{{\em Phys. Rev. D}
  {\bfseries 98} (2018) 092014}.

\bibitem{Tripathee:2017ybi}
A.~Tripathee, W.~Xue, A.~Larkoski, S.~Marzani, and J.~Thaler, ``{Jet
  Substructure Studies with CMS Open Data}'',
  \href{http://dx.doi.org/10.1103/PhysRevD.96.074003}{{\em Phys. Rev. D}
  {\bfseries 96} (2017) 074003}.

\bibitem{STAR2020}
{\bfseries STAR} Collaboration, J.~Adam {\em et~al.}, ``{Measurement of groomed
  jet substructure observables in p+p collisions at $\sqrt {s}$ =200 GeV with
  STAR}'', \href{http://dx.doi.org/10.1016/j.physletb.2020.135846}{{\em Phys.
  Lett. B} {\bfseries 811} (2020) 135846},
  \href{http://arxiv.org/abs/2003.02114}{{\ttfamily arXiv:2003.02114
  [hep-ex]}}.

\bibitem{Kang:2019prh}
Z.-B. Kang, K.~Lee, X.~Liu, D.~Neill, and F.~Ringer, ``{The soft drop groomed
  jet radius at NLL}'', \href{http://dx.doi.org/10.1007/JHEP02(2020)054}{{\em
  J. High Energ. Phys.} {\bfseries 02} (2020) 054},
  \href{http://arxiv.org/abs/1908.01783}{{\ttfamily arXiv:1908.01783
  [hep-ph]}}.

\bibitem{Ringer:2019rfk}
F.~Ringer, B.-W. Xiao, and F.~Yuan, ``{Can we observe jet $P_T$-broadening in
  heavy-ion collisions at the LHC?}'',
  \href{http://dx.doi.org/10.1016/j.physletb.2020.135634}{{\em Phys. Lett. B}
  {\bfseries 808} (2020) 135634},
  \href{http://arxiv.org/abs/1907.12541}{{\ttfamily arXiv:1907.12541
  [hep-ph]}}.

\bibitem{CasalderreySolana:2012ef}
J.~Casalderrey-Solana, Y.~Mehtar-Tani, C.~A. Salgado, and K.~Tywoniuk, ``{New
  picture of jet quenching dictated by color coherence}'',
  \href{http://dx.doi.org/10.1016/j.physletb.2013.07.046}{{\em Phys. Lett. B}
  {\bfseries 725} (2013) 357--360}.

\bibitem{Chien:2016led}
Y.-T. Chien and I.~Vitev, ``{Probing the Hardest Branching within Jets in
  Heavy-Ion Collisions}'',
  \href{http://dx.doi.org/10.1103/PhysRevLett.119.112301}{{\em Phys. Rev.
  Lett.} {\bfseries 119} (2017) 112301}.

\bibitem{Caucal:2019uvr}
P.~Caucal, E.~Iancu, and G.~Soyez, ``{Deciphering the $z_g$ distribution in
  ultrarelativistic heavy ion collisions}'',
  \href{http://dx.doi.org/10.1007/JHEP10(2019)273}{{\em J. High Energ. Phys.}
  {\bfseries 10} (2019) 273}.

\bibitem{Chang:2019nrx}
N.-B. Chang, S.~Cao, and G.-Y. Qin, ``Probing medium-induced jet splitting and
  energy loss in heavy-ion collisions'',
  \href{http://dx.doi.org/10.1016/j.physletb.2018.04.019}{{\em Physics Letters
  B} {\bfseries 781} (2018) 423–432}.

\bibitem{Casalderrey-Solana:2019ubu}
J.~Casalderrey-Solana, G.~Milhano, D.~Pablos, and K.~Rajagopal, ``{Modification
  of Jet Substructure in Heavy Ion Collisions as a Probe of the Resolution
  Length of Quark-Gluon Plasma}'',
  \href{http://dx.doi.org/10.1007/JHEP01(2020)044}{{\em J. High Energ. Phys.}
  {\bfseries 01} (2020) 044}.

\bibitem{mulligan2020identifying}
J.~Mulligan and M.~Ploskon, ``{Identifying groomed jet splittings in heavy-ion
  collisions}'', \href{http://dx.doi.org/10.1103/PhysRevC.102.044913}{{\em
  Phys. Rev. C} {\bfseries 102} (2020) 044913},
  \href{http://arxiv.org/abs/2006.01812}{{\ttfamily arXiv:2006.01812
  [hep-ph]}}.

\bibitem{aliceDetector}
{\bfseries ALICE} Collaboration, K.~Aamodt {\em et~al.}, ``{The ALICE
  experiment at the CERN LHC}'',
  \href{http://dx.doi.org/10.1088/1748-0221/3/08/S08002}{{\em JINST} {\bfseries
  3} (2008) S08002}.

\bibitem{Abelev:2014ffa}
{\bfseries ALICE} Collaboration, ``{Performance of the ALICE Experiment at the
  CERN LHC}'', \href{http://dx.doi.org/10.1142/S0217751X14300440}{{\em Int. J.
  Mod. Phys. A} {\bfseries 29} (2014) 1430044}.

\bibitem{Abbas:2013taa}
{\bfseries ALICE} Collaboration, ``{Performance of the ALICE VZERO system}'',
  \href{http://dx.doi.org/10.1088/1748-0221/8/10/P10016}{{\em JINST} {\bfseries
  8} (2013) P10016}.

\bibitem{centrality276}
{\bfseries ALICE} Collaboration, ``{Centrality determination of \PbPb{}
  collisions at $\sqrts=2.76$ TeV with ALICE}'',
  \href{http://dx.doi.org/10.1103/PhysRevC.88.044909}{{\em Phys. Rev. C}
  {\bfseries 88} (2013) 044909}.

\bibitem{centrality502}
{\bfseries ALICE} Collaboration, ``{Centrality determination in heavy-ion
  collisions}.''
\newblock \url{http://cds.cern.ch/record/2636623}.

\bibitem{ppXsec}
{\bfseries ALICE} Collaboration, ``{ALICE 2017 luminosity determination for pp
  collisions at $\sqrt{s}=$5 TeV}'',. \url{http://cds.cern.ch/record/2648933}.

\bibitem{Alme_2010}
J.~Alme and et~al., ``{The ALICE TPC, a large 3-dimensional tracking device
  with fast readout for ultra-high multiplicity events}'',
  \href{http://dx.doi.org/10.1016/j.nima.2010.04.042}{{\em Nucl. Instrum. Meth.
  A: Accelerators, Spectrometers, Detectors and Associated Equipment}
  {\bfseries 622} (2010) 316–367}.

\bibitem{Aamodt:2010aa}
{\bfseries ALICE} Collaboration, ``{Alignment of the ALICE Inner Tracking
  System with cosmic-ray tracks}'',
  \href{http://dx.doi.org/10.1088/1748-0221/5/03/P03003}{{\em JINST} {\bfseries
  5} (2010) P03003}.

\bibitem{Chang:2013rca}
H.-M. Chang, M.~Procura, J.~Thaler, and W.~J. Waalewijn, ``{Calculating
  Track-Based Observables for the LHC}'',
  \href{http://dx.doi.org/10.1103/PhysRevLett.111.102002}{{\em Phys. Rev.
  Lett.} {\bfseries 111} (Sep, 2013) 102002},
  \href{http://arxiv.org/abs/1303.6637}{{\ttfamily arXiv:1303.6637 [hep-ph]}}.

\bibitem{Chen:2020vvp}
H.~Chen, I.~Moult, X.~Zhang, and H.~X. Zhu, ``{Rethinking jets with energy
  correlators: Tracks, resummation, and analytic continuation}'',
  \href{http://dx.doi.org/10.1103/PhysRevD.102.054012}{{\em Phys. Rev. D}
  {\bfseries 102} no.~5, (Sep, 2020) 054012},
  \href{http://arxiv.org/abs/2004.11381}{{\ttfamily arXiv:2004.11381
  [hep-ph]}}.

\bibitem{Chien:2020hzh}
Y.-T. Chien, R.~Rahn, S.~Schrijnder~van Velzen, D.~Y. Shao, W.~J. Waalewijn,
  and B.~Wu, ``{Recoil-free azimuthal angle for precision boson-jet
  correlation}'', \href{http://dx.doi.org/10.1016/j.physletb.2021.136124}{{\em
  Phys. Lett. B} {\bfseries 815} (2021) 136124},
  \href{http://arxiv.org/abs/2005.12279}{{\ttfamily arXiv:2005.12279
  [hep-ph]}}.

\bibitem{ATLAS:2019mgf}
{\bfseries ATLAS} Collaboration, G.~Aad {\em et~al.}, ``{Measurement of
  soft-drop jet observables in $pp$ collisions with the ATLAS detector at
  $\sqrt {s}$ =13 TeV}'',
  \href{http://dx.doi.org/10.1103/PhysRevD.101.052007}{{\em Phys. Rev. D}
  {\bfseries 101} no.~5, (2020) 052007},
  \href{http://arxiv.org/abs/1912.09837}{{\ttfamily arXiv:1912.09837
  [hep-ex]}}.

\bibitem{THE_PUBLIC_NOTE}
{\bfseries ALICE} Collaboration, ``{Supplemental material: Measurements of the
  groomed jet radius and groomed momentum fraction in pp and Pb--Pb collisions
  at $\sqrts = 5.02$ TeV}.''
\newblock \url{https://cds.cern.ch/record/2725572}.

\bibitem{Cacciari:2011ma}
M.~Cacciari, G.~P. Salam, and G.~Soyez, ``{FastJet User Manual}'',
  \href{http://dx.doi.org/10.1140/epjc/s10052-012-1896-2}{{\em Eur. Phys. J. C}
  {\bfseries 72} (2012) 1896}.

\bibitem{antikt}
M.~Cacciari, G.~P. Salam, and G.~Soyez, ``{The anti-$k_t$ jet clustering
  algorithm}'', \href{http://dx.doi.org/10.1088/1126-6708/2008/04/063}{{\em J.
  High Energ. Phys.} {\bfseries 04} (2008) 063}.

\bibitem{catchment}
M.~Cacciari, G.~P. Salam, and G.~Soyez, ``{The Catchment Area of Jets}'',
  \href{http://dx.doi.org/10.1088/1126-6708/2008/04/005}{{\em J. High Energ.
  Phys.} {\bfseries 04} (2008) 005}.

\bibitem{Abelev2012}
{\bfseries ALICE} Collaboration, ``{Measurement of event background
  fluctuations for charged particle jet reconstruction in Pb--Pb collisions at
  $\sqrts=2.76$ TeV}'', \href{http://dx.doi.org/10.1007/JHEP03(2012)053}{{\em
  J. High Energ. Phys.} {\bfseries 2012} (2012) 53}.

\bibitem{Berta:2014eza}
P.~Berta, M.~Spousta, D.~W. Miller, and R.~Leitner, ``{Particle-level pileup
  subtraction for jets and jet shapes}'',
  \href{http://dx.doi.org/10.1007/JHEP06(2014)092}{{\em J. High Energ. Phys.}
  {\bfseries 06} (2014) 092}.

\bibitem{Berta:2019hnj}
P.~Berta, L.~Masetti, D.~Miller, and M.~Spousta, ``{Pileup and Underlying Event
  Mitigation with Iterative Constituent Subtraction}'',
  \href{http://dx.doi.org/10.1007/JHEP08(2019)175}{{\em J. High Energ. Phys.}
  {\bfseries 08} (2019) 175}.

\bibitem{pythia}
T.~Sjostrand, S.~Ask, J.~R. Christiansen, R.~Corke, N.~Desai, P.~Ilten,
  S.~Mrenna, S.~Prestel, C.~O. Rasmussen, and P.~Z. Skands, ``{An introduction
  to PYTHIA 8.2}'', \href{http://dx.doi.org/10.1016/j.cpc.2015.01.024}{{\em
  Comput. Phys. Commun.} {\bfseries 191} (2015) 159--177}.

\bibitem{Brun:1119728}
R.~Brun, F.~Bruyant, M.~Maire, A.~C. McPherson, and P.~Zanarini, {\em {GEANT 3:
  user's guide Geant 3.10, Geant 3.11; rev. version}}.
\newblock CERN, Geneva, 1987.
\newblock \url{https://cds.cern.ch/record/1119728}.

\bibitem{DAgostini}
G.~D'Agostini, ``A multidimensional unfolding method based on bayes' theorem'',
  \href{http://dx.doi.org/10.1016/0168-9002(95)00274-X}{{\em Nucl. Instrum.
  Meth. A: Accelerators, Spectrometers, Detectors and Associated Equipment}
  {\bfseries 362} (1995) 487 -- 498}.

\bibitem{roounfold}
``{RooUnfold}.''
\newblock \url{http://hepunx.rl.ac.uk/~adye/software/unfold/RooUnfold.html}.
  Access date: May 31 2020.

\bibitem{Bellm:2015jjp}
J.~Bellm {\em et~al.}, ``{Herwig 7.0/Herwig++ 3.0 release note}'',
  \href{http://dx.doi.org/10.1140/epjc/s10052-016-4018-8}{{\em Eur. Phys. J. C}
  {\bfseries 76} no.~4, (Apr, 2016) 196},
  \href{http://arxiv.org/abs/1512.01178}{{\ttfamily arXiv:1512.01178
  [hep-ph]}}.

\bibitem{Zapp:2013vla}
K.~C. Zapp, ``{JEWEL 2.0.0: directions for use}'',
  \href{http://dx.doi.org/10.1140/epjc/s10052-014-2762-1}{{\em Eur. Phys. J. C}
  {\bfseries 74} no.~2, (2014) 2762},
  \href{http://arxiv.org/abs/1311.0048}{{\ttfamily arXiv:1311.0048 [hep-ph]}}.

\bibitem{Putschke:2019yrg}
{\bfseries JETSCAPE} Collaboration, J.~Putschke {\em et~al.}, ``{The JETSCAPE
  framework}'', \href{http://arxiv.org/abs/1903.07706}{{\ttfamily
  arXiv:1903.07706 [nucl-th]}}.

\bibitem{Zapp:2012ak}
K.~C. Zapp, F.~Krauss, and U.~A. Wiedemann, ``{A perturbative framework for jet
  quenching}'', \href{http://dx.doi.org/10.1007/JHEP03(2013)080}{{\em J. High
  Energ. Phys.} {\bfseries 03} (2013) 080},
  \href{http://arxiv.org/abs/1212.1599}{{\ttfamily arXiv:1212.1599 [hep-ph]}}.

\bibitem{Caucal_2018}
P.~Caucal, E.~Iancu, A.~H. Mueller, and G.~Soyez, ``{Vacuum-like jet
  fragmentation in a dense QCD medium}'',
  \href{http://dx.doi.org/10.1103/PhysRevLett.120.232001}{{\em Phys. Rev.
  Lett.} {\bfseries 120} (2018) 232001},
  \href{http://arxiv.org/abs/1801.09703}{{\ttfamily arXiv:1801.09703
  [hep-ph]}}.

\bibitem{HybridModel}
J.~Casalderrey-Solana, D.~C. Gulhan, J.~G. Milhano, D.~Pablos, and
  K.~Rajagopal, ``{A Hybrid Strong/Weak Coupling Approach to Jet Quenching}'',
  \href{http://dx.doi.org/10.1007/JHEP09(2015)175}{{\em J. High Energ. Phys.}
  {\bfseries 10} (2014) 019}, \href{http://arxiv.org/abs/1405.3864}{{\ttfamily
  arXiv:1405.3864 [hep-ph]}}. [Erratum: \textit{JHEP} \textbf{09} (2015), 175].

\bibitem{HybridModelResolution}
Z.~Hulcher, D.~Pablos, and K.~Rajagopal, ``{Resolution Effects in the Hybrid
  Strong/Weak Coupling Model}'',
  \href{http://dx.doi.org/10.1007/JHEP03(2018)010}{{\em J. High Energ. Phys.}
  {\bfseries 03} (2018) 010}, \href{http://arxiv.org/abs/1707.05245}{{\ttfamily
  arXiv:1707.05245 [hep-ph]}}.

\bibitem{Majumder_2013}
A.~Majumder, ``{Incorporating Space-Time Within Medium-Modified Jet Event
  Generators}'', \href{http://dx.doi.org/10.1103/PhysRevC.88.014909}{{\em Phys.
  Rev. C} {\bfseries 88} (2013) 014909},
  \href{http://arxiv.org/abs/1301.5323}{{\ttfamily arXiv:1301.5323 [nucl-th]}}.

\bibitem{LBT}
Y.~He, T.~Luo, X.-N. Wang, and Y.~Zhu, ``{Linear Boltzmann Transport for Jet
  Propagation in the Quark-Gluon Plasma: Elastic Processes and Medium
  Recoil}'', \href{http://dx.doi.org/10.1103/PhysRevC.91.054908}{{\em Phys.
  Rev. C} {\bfseries 91} (2015) 054908},
  \href{http://arxiv.org/abs/1503.03313}{{\ttfamily arXiv:1503.03313
  [nucl-th]}}. [Erratum: \textit{Phys.Rev.C} \textbf{97} (2018), 019902].

\bibitem{KunnawalkamElayavalli:2017hxo}
R.~Kunnawalkam~Elayavalli and K.~C. Zapp, ``{Medium response in JEWEL and its
  impact on jet shape observables in heavy ion collisions}'',
  \href{http://dx.doi.org/10.1007/JHEP07(2017)141}{{\em JHEP} {\bfseries 07}
  (2017) 141}, \href{http://arxiv.org/abs/1707.01539}{{\ttfamily
  arXiv:1707.01539 [hep-ph]}}.

\bibitem{Qiu:2019sfj}
J.-W. Qiu, F.~Ringer, N.~Sato, and P.~Zurita, ``{Factorization of jet cross
  sections in heavy-ion collisions}'',
  \href{http://dx.doi.org/10.1103/PhysRevLett.122.252301}{{\em Phys. Rev.
  Lett.} {\bfseries 122} (2019) 252301}.

\end{thebibliography}\endgroup

%%%%%%%%%%%%%%%%%%%%%%%%%%%%%%%%
% Appendices: yours (if any) + authorlist
%%%%%%%%%%%%%%%%%%%%%%%%%%%%%%%%
\newpage
\appendix

%
%\input{} % put your appendices here (if any)
%

%%%%% Authorlist - please do not touch: handled by EB chairs 
\section{The ALICE Collaboration}
\label{app:collab}
% ALICE Collaboration author list for 2021-07-15
\small
\begin{flushleft}

S.~Acharya$^{\rm 143}$, 
D.~Adamov\'{a}$^{\rm 98}$, 
A.~Adler$^{\rm 76}$, 
G.~Aglieri Rinella$^{\rm 35}$, 
M.~Agnello$^{\rm 31}$, 
N.~Agrawal$^{\rm 55}$, 
Z.~Ahammed$^{\rm 143}$, 
S.~Ahmad$^{\rm 16}$, 
S.U.~Ahn$^{\rm 78}$, 
I.~Ahuja$^{\rm 39}$
% ...

\section*{Affiliation notes}

$^{\rm I}$ Deceased\\
$^{\rm II}$ Also at: Italian National Agency for New Technologies, Energy and Sustainable Economic Development (ENEA), Bologna, Italy\\
$^{\rm III}$ Also at: Dipartimento DET del Politecnico di Torino, Turin, Italy\\
$^{\rm IV}$ Also at: M.V. Lomonosov Moscow State University, D.V. Skobeltsyn Institute of Nuclear, Physics, Moscow, Russia\\
$^{\rm V}$ Also at: Department of Applied Physics, Aligarh Muslim University, Aligarh, India
\\
$^{\rm VI}$ Also at: Institute of Theoretical Physics, University of Wroclaw, Poland\\
$^{\rm VII}$ Also at: University of Kansas, Lawrence, Kansas, United States\\

\section*{Collaboration Institutes}

$^{1}$ A.I. Alikhanyan National Science Laboratory (Yerevan Physics Institute) Foundation, Yerevan, Armenia\\
$^{2}$ AGH University of Science and Technology, Cracow, Poland\\
$^{3}$ Bogolyubov Institute for Theoretical Physics, National Academy of Sciences of Ukraine, Kiev, Ukraine\\
$^{4}$ Bose Institute, Department of Physics  and Centre for Astroparticle Physics and Space Science (CAPSS), Kolkata, India\\
$^{5}$ Budker Institute for Nuclear Physics, Novosibirsk, Russia\\
$^{6}$ California Polytechnic State University, San Luis Obispo, California, United States\\
$^{7}$ Central China Normal University, Wuhan, China\\
$^{8}$ Centro de Aplicaciones Tecnol\'{o}gicas y Desarrollo Nuclear (CEADEN), Havana, Cuba\\
$^{9}$ Centro de Investigaci\'{o}n y de Estudios Avanzados (CINVESTAV), Mexico City and M\'{e}rida, Mexico\\
$^{10}$ Chicago State University, Chicago, Illinois, United States\\
$^{11}$ China Institute of Atomic Energy, Beijing, China\\
$^{12}$ Chungbuk National University, Cheongju, Republic of Korea\\
$^{13}$ Comenius University Bratislava, Faculty of Mathematics, Physics and Informatics, Bratislava, Slovakia\\
$^{14}$ COMSATS University Islamabad, Islamabad, Pakistan\\
$^{15}$ Creighton University, Omaha, Nebraska, United States\\
$^{16}$ Department of Physics, Aligarh Muslim University, Aligarh, India\\
$^{17}$ Department of Physics, Pusan National University, Pusan, Republic of Korea\\
$^{18}$ Department of Physics, Sejong University, Seoul, Republic of Korea\\
$^{19}$ Department of Physics, University of California, Berkeley, California, United States\\
$^{20}$ Department of Physics, University of Oslo, Oslo, Norway\\
$^{21}$ Department of Physics and Technology, University of Bergen, Bergen, Norway\\
$^{22}$ Dipartimento di Fisica dell'Universit\`{a} 'La Sapienza' and Sezione INFN, Rome, Italy\\
$^{23}$ Dipartimento di Fisica dell'Universit\`{a} and Sezione INFN, Cagliari, Italy\\
$^{24}$ Dipartimento di Fisica dell'Universit\`{a} and Sezione INFN, Trieste, Italy\\
$^{25}$ Dipartimento di Fisica dell'Universit\`{a} and Sezione INFN, Turin, Italy\\
$^{26}$ Dipartimento di Fisica e Astronomia dell'Universit\`{a} and Sezione INFN, Bologna, Italy\\
$^{27}$ Dipartimento di Fisica e Astronomia dell'Universit\`{a} and Sezione INFN, Catania, Italy\\
$^{28}$ Dipartimento di Fisica e Astronomia dell'Universit\`{a} and Sezione INFN, Padova, Italy\\
$^{29}$ Dipartimento di Fisica e Nucleare e Teorica, Universit\`{a} di Pavia, Pavia, Italy\\
$^{30}$ Dipartimento di Fisica `E.R.~Caianiello' dell'Universit\`{a} and Gruppo Collegato INFN, Salerno, Italy\\
$^{31}$ Dipartimento DISAT del Politecnico and Sezione INFN, Turin, Italy\\
$^{32}$ Dipartimento di Scienze e Innovazione Tecnologica dell'Universit\`{a} del Piemonte Orientale and INFN Sezione di Torino, Alessandria, Italy\\
$^{33}$ Dipartimento di Scienze MIFT, Universit\`{a} di Messina, Messina, Italy\\
$^{34}$ Dipartimento Interateneo di Fisica `M.~Merlin' and Sezione INFN, Bari, Italy\\
$^{35}$ European Organization for Nuclear Research (CERN), Geneva, Switzerland\\
$^{36}$ Faculty of Electrical Engineering, Mechanical Engineering and Naval Architecture, University of Split, Split, Croatia\\
$^{37}$ Faculty of Engineering and Science, Western Norway University of Applied Sciences, Bergen, Norway\\
$^{38}$ Faculty of Nuclear Sciences and Physical Engineering, Czech Technical University in Prague, Prague, Czech Republic\\
$^{39}$ Faculty of Science, P.J.~\v{S}af\'{a}rik University, Ko\v{s}ice, Slovakia\\
$^{40}$ Frankfurt Institute for Advanced Studies, Johann Wolfgang Goethe-Universit\"{a}t Frankfurt, Frankfurt, Germany\\
$^{41}$ Fudan University, Shanghai, China\\
$^{42}$ Gangneung-Wonju National University, Gangneung, Republic of Korea\\
$^{43}$ Gauhati University, Department of Physics, Guwahati, India\\
$^{44}$ Helmholtz-Institut f\"{u}r Strahlen- und Kernphysik, Rheinische Friedrich-Wilhelms-Universit\"{a}t Bonn, Bonn, Germany\\
$^{45}$ Helsinki Institute of Physics (HIP), Helsinki, Finland\\
$^{46}$ High Energy Physics Group,  Universidad Aut\'{o}noma de Puebla, Puebla, Mexico\\
$^{47}$ Hiroshima University, Hiroshima, Japan\\
$^{48}$ Hochschule Worms, Zentrum  f\"{u}r Technologietransfer und Telekommunikation (ZTT), Worms, Germany\\
$^{49}$ Horia Hulubei National Institute of Physics and Nuclear Engineering, Bucharest, Romania\\
$^{50}$ Indian Institute of Technology Bombay (IIT), Mumbai, India\\
$^{51}$ Indian Institute of Technology Indore, Indore, India\\
$^{52}$ Indonesian Institute of Sciences, Jakarta, Indonesia\\
$^{53}$ INFN, Laboratori Nazionali di Frascati, Frascati, Italy\\
$^{54}$ INFN, Sezione di Bari, Bari, Italy\\
$^{55}$ INFN, Sezione di Bologna, Bologna, Italy\\
$^{56}$ INFN, Sezione di Cagliari, Cagliari, Italy\\
$^{57}$ INFN, Sezione di Catania, Catania, Italy\\
$^{58}$ INFN, Sezione di Padova, Padova, Italy\\
$^{59}$ INFN, Sezione di Pavia, Pavia, Italy\\
$^{60}$ INFN, Sezione di Roma, Rome, Italy\\
$^{61}$ INFN, Sezione di Torino, Turin, Italy\\
$^{62}$ INFN, Sezione di Trieste, Trieste, Italy\\
$^{63}$ Inha University, Incheon, Republic of Korea\\
$^{64}$ Institute for Gravitational and Subatomic Physics (GRASP), Utrecht University/Nikhef, Utrecht, Netherlands\\
$^{65}$ Institute for Nuclear Research, Academy of Sciences, Moscow, Russia\\
$^{66}$ Institute of Experimental Physics, Slovak Academy of Sciences, Ko\v{s}ice, Slovakia\\
$^{67}$ Institute of Physics, Homi Bhabha National Institute, Bhubaneswar, India\\
$^{68}$ Institute of Physics of the Czech Academy of Sciences, Prague, Czech Republic\\
$^{69}$ Institute of Space Science (ISS), Bucharest, Romania\\
$^{70}$ Institut f\"{u}r Kernphysik, Johann Wolfgang Goethe-Universit\"{a}t Frankfurt, Frankfurt, Germany\\
$^{71}$ Instituto de Ciencias Nucleares, Universidad Nacional Aut\'{o}noma de M\'{e}xico, Mexico City, Mexico\\
$^{72}$ Instituto de F\'{i}sica, Universidade Federal do Rio Grande do Sul (UFRGS), Porto Alegre, Brazil\\
$^{73}$ Instituto de F\'{\i}sica, Universidad Nacional Aut\'{o}noma de M\'{e}xico, Mexico City, Mexico\\
$^{74}$ iThemba LABS, National Research Foundation, Somerset West, South Africa\\
$^{75}$ Jeonbuk National University, Jeonju, Republic of Korea\\
$^{76}$ Johann-Wolfgang-Goethe Universit\"{a}t Frankfurt Institut f\"{u}r Informatik, Fachbereich Informatik und Mathematik, Frankfurt, Germany\\
$^{77}$ Joint Institute for Nuclear Research (JINR), Dubna, Russia\\
$^{78}$ Korea Institute of Science and Technology Information, Daejeon, Republic of Korea\\
$^{79}$ KTO Karatay University, Konya, Turkey\\
$^{80}$ Laboratoire de Physique des 2 Infinis, Ir\`{e}ne Joliot-Curie, Orsay, France\\
$^{81}$ Laboratoire de Physique Subatomique et de Cosmologie, Universit\'{e} Grenoble-Alpes, CNRS-IN2P3, Grenoble, France\\
$^{82}$ Lawrence Berkeley National Laboratory, Berkeley, California, United States\\
$^{83}$ Lund University Department of Physics, Division of Particle Physics, Lund, Sweden\\
$^{84}$ Moscow Institute for Physics and Technology, Moscow, Russia\\
$^{85}$ Nagasaki Institute of Applied Science, Nagasaki, Japan\\
$^{86}$ Nara Women{'}s University (NWU), Nara, Japan\\
$^{87}$ National and Kapodistrian University of Athens, School of Science, Department of Physics , Athens, Greece\\
$^{88}$ National Centre for Nuclear Research, Warsaw, Poland\\
$^{89}$ National Institute of Science Education and Research, Homi Bhabha National Institute, Jatni, India\\
$^{90}$ National Nuclear Research Center, Baku, Azerbaijan\\
$^{91}$ National Research Centre Kurchatov Institute, Moscow, Russia\\
$^{92}$ Niels Bohr Institute, University of Copenhagen, Copenhagen, Denmark\\
$^{93}$ Nikhef, National institute for subatomic physics, Amsterdam, Netherlands\\
$^{94}$ NRC Kurchatov Institute IHEP, Protvino, Russia\\
$^{95}$ NRC \guillemotleft Kurchatov\guillemotright  Institute - ITEP, Moscow, Russia\\
$^{96}$ NRNU Moscow Engineering Physics Institute, Moscow, Russia\\
$^{97}$ Nuclear Physics Group, STFC Daresbury Laboratory, Daresbury, United Kingdom\\
$^{98}$ Nuclear Physics Institute of the Czech Academy of Sciences, \v{R}e\v{z} u Prahy, Czech Republic\\
$^{99}$ Oak Ridge National Laboratory, Oak Ridge, Tennessee, United States\\
$^{100}$ Ohio State University, Columbus, Ohio, United States\\
$^{101}$ Petersburg Nuclear Physics Institute, Gatchina, Russia\\
$^{102}$ Physics department, Faculty of science, University of Zagreb, Zagreb, Croatia\\
$^{103}$ Physics Department, Panjab University, Chandigarh, India\\
$^{104}$ Physics Department, University of Jammu, Jammu, India\\
$^{105}$ Physics Department, University of Rajasthan, Jaipur, India\\
$^{106}$ Physikalisches Institut, Eberhard-Karls-Universit\"{a}t T\"{u}bingen, T\"{u}bingen, Germany\\
$^{107}$ Physikalisches Institut, Ruprecht-Karls-Universit\"{a}t Heidelberg, Heidelberg, Germany\\
$^{108}$ Physik Department, Technische Universit\"{a}t M\"{u}nchen, Munich, Germany\\
$^{109}$ Politecnico di Bari and Sezione INFN, Bari, Italy\\
$^{110}$ Research Division and ExtreMe Matter Institute EMMI, GSI Helmholtzzentrum f\"ur Schwerionenforschung GmbH, Darmstadt, Germany\\
$^{111}$ Russian Federal Nuclear Center (VNIIEF), Sarov, Russia\\
$^{112}$ Saha Institute of Nuclear Physics, Homi Bhabha National Institute, Kolkata, India\\
$^{113}$ School of Physics and Astronomy, University of Birmingham, Birmingham, United Kingdom\\
$^{114}$ Secci\'{o}n F\'{\i}sica, Departamento de Ciencias, Pontificia Universidad Cat\'{o}lica del Per\'{u}, Lima, Peru\\
$^{115}$ St. Petersburg State University, St. Petersburg, Russia\\
$^{116}$ Stefan Meyer Institut f\"{u}r Subatomare Physik (SMI), Vienna, Austria\\
$^{117}$ SUBATECH, IMT Atlantique, Universit\'{e} de Nantes, CNRS-IN2P3, Nantes, France\\
$^{118}$ Suranaree University of Technology, Nakhon Ratchasima, Thailand\\
$^{119}$ Technical University of Ko\v{s}ice, Ko\v{s}ice, Slovakia\\
$^{120}$ The Henryk Niewodniczanski Institute of Nuclear Physics, Polish Academy of Sciences, Cracow, Poland\\
$^{121}$ The University of Texas at Austin, Austin, Texas, United States\\
$^{122}$ Universidad Aut\'{o}noma de Sinaloa, Culiac\'{a}n, Mexico\\
$^{123}$ Universidade de S\~{a}o Paulo (USP), S\~{a}o Paulo, Brazil\\
$^{124}$ Universidade Estadual de Campinas (UNICAMP), Campinas, Brazil\\
$^{125}$ Universidade Federal do ABC, Santo Andre, Brazil\\
$^{126}$ University of Cape Town, Cape Town, South Africa\\
$^{127}$ University of Houston, Houston, Texas, United States\\
$^{128}$ University of Jyv\"{a}skyl\"{a}, Jyv\"{a}skyl\"{a}, Finland\\
$^{129}$ University of Kansas, Lawrence, Kansas, United States\\
$^{130}$ University of Liverpool, Liverpool, United Kingdom\\
$^{131}$ University of Science and Technology of China, Hefei, China\\
$^{132}$ University of South-Eastern Norway, Tonsberg, Norway\\
$^{133}$ University of Tennessee, Knoxville, Tennessee, United States\\
$^{134}$ University of the Witwatersrand, Johannesburg, South Africa\\
$^{135}$ University of Tokyo, Tokyo, Japan\\
$^{136}$ University of Tsukuba, Tsukuba, Japan\\
$^{137}$ Universit\'{e} Clermont Auvergne, CNRS/IN2P3, LPC, Clermont-Ferrand, France\\
$^{138}$ Universit\'{e} de Lyon, CNRS/IN2P3, Institut de Physique des 2 Infinis de Lyon , Lyon, France\\
$^{139}$ Universit\'{e} de Strasbourg, CNRS, IPHC UMR 7178, F-67000 Strasbourg, France, Strasbourg, France\\
$^{140}$ Universit\'{e} Paris-Saclay Centre d'Etudes de Saclay (CEA), IRFU, D\'{e}partment de Physique Nucl\'{e}aire (DPhN), Saclay, France\\
$^{141}$ Universit\`{a} degli Studi di Foggia, Foggia, Italy\\
$^{142}$ Universit\`{a} di Brescia, Brescia, Italy\\
$^{143}$ Variable Energy Cyclotron Centre, Homi Bhabha National Institute, Kolkata, India\\
$^{144}$ Warsaw University of Technology, Warsaw, Poland\\
$^{145}$ Wayne State University, Detroit, Michigan, United States\\
$^{146}$ Westf\"{a}lische Wilhelms-Universit\"{a}t M\"{u}nster, Institut f\"{u}r Kernphysik, M\"{u}nster, Germany\\
$^{147}$ Wigner Research Centre for Physics, Budapest, Hungary\\
$^{148}$ Yale University, New Haven, Connecticut, United States\\
$^{149}$ Yonsei University, Seoul, Republic of Korea\\

\end{flushleft}   
\end{document}